\documentclass[11pt,a4paper]{article}
\pdfoutput=1 

\usepackage{jinstpub} 

\title{A marine radioisotope gamma-ray spectrum analysis method based on Geant4 simulation and MLP neural network}

\author[a]{Wenhan Dai,}
\author[a]{Zhi Zeng,}
\author[a]{Daowei Dou,}
\author[a,1]{Hao Ma,\note{Corresponding author.}}
\author[a,b]{Jianping Chen,}
\author[a]{Junli Li,}
\author[a]{and Hui Zhang}

\affiliation[a]{Key Laboratory of Particle and Radiation Imaging (Ministry of Education) and Department of Engineering Physics, Tsinghua University,\\Beijing 100084, China}
\affiliation[b]{College of Nuclear Science and Technology, Beijing Normal University,\\Beijing 100875, China}

\emailAdd{ mahao@tsinghua.edu.cn}

\abstract{The monitoring of marine $^{\text{137}}$Cs using a scintillation detector relies on the spectrum analysis method to extract the $^{\text{137}}$Cs concentration. And when in a poor statistic situation, the calculation result of the traditional net peak area (NPA) method has a large uncertainty. We present a new machine learning based method to better analyze the gamma-ray spectrum with low $^{\text{137}}$Cs concentration. We apply multilayer perceptron (MLP) to analyze the 662 keV full energy peak of $^{\text{137}}$Cs in the seawater spectrum. And the MLP can be trained with a few measured background spectrums by combining the simulated $^{\text{137}}$Cs spectrums with measured background spectrums. Thus, it can save the time of preparing and measuring the standard samples for generating the training dataset. To validate the MLP-based method, we use Geant4 and background gamma-ray spectrums measured by a seaborne monitoring device to generate an independent test dataset to test the result by our MLP-based method and the traditional NPA method. We find that the MLP-based method achieves a root mean squared error of 0.159, 2.3 times lower than that of the traditional net peak area method, indicating the MLP-based method improves the precision of $^{\text{137}}$Cs concentration calculation}


\begin{document}
\maketitle
\flushbottom

\section{Introduction}
\label{sec:intro}

In Fukushima nuclear accident in 2011, radionuclides were released to the oceanic environment via atmospheric fallout and wastewater discharge, with the $^{\text{137}}$Cs activity of 3.1-3.6 PBq \cite{bib:1, bib:2}. Due to the long half-life (30.17 y) and transportation to east by currents in the Pacific Ocean, it is important and necessary to implement extensive and long-term seaborne monitoring of $^{\text{137}}$Cs concentration in seawater.

\begin{figure*}[!htb]
\includegraphics
  [width=0.9\hsize]
{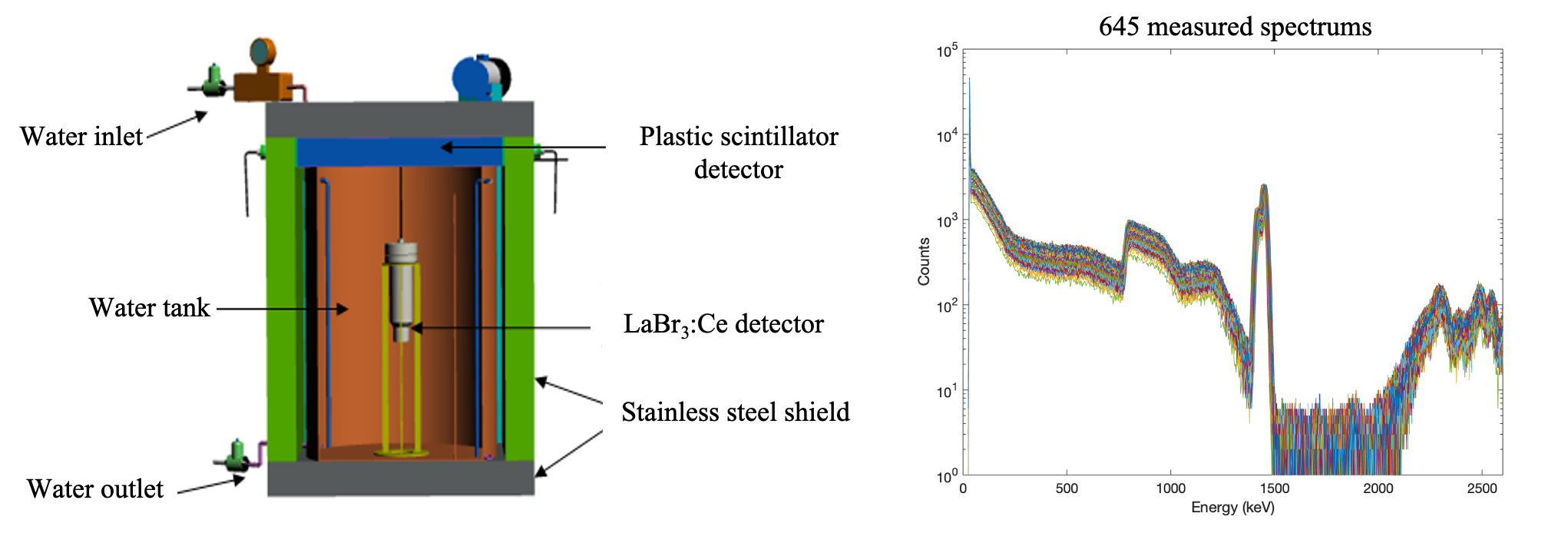}
\caption{Sketch of the marine radioactive isotope monitoring device (left) and 645 spectrums measured in the sea trail (right).}
\label{fig:figure-1}
\end{figure*}

In our previous work, a seaborne monitoring device has been developed to monitor the radioactive isotopes \cite{bib:3, bib:4}. As shown in Fig.~\ref{fig:figure-1} (left), the device is equipped with a 2×2 inches LaBr$_{\text{3}}$ detector and shielded by stainless steel. When monitoring, the seawater is pumped into the device chamber. And after the chamber is filled, the seawater is measured by a LaBr$_{\text{3}}$ detector.

The net-peak area method is commonly used in the gamma-ray spectrum analysis, and the $^{\text{137}}$Cs concentration is calculated by the detection efficiency and the net peak counts of the 662 keV full energy peak of $^{\text{137}}$Cs. However, during the environmental radioactive level measurement, we need to measure $^{\text{137}}$Cs concentration near the detection limit, and the calculation result of the traditional method has a large uncertainty because of the poor statistic, thereby may lead to an omission or a false alarm.

Neural network is a powerful tool to automatically analyze a gamma-ray spectrum by training the network with known spectrums. Due to the advantages of self-adapting and time-saving, it has been widely applied in gamma-ray spectrometry, such as activity estimation \cite{bib:5, bib:14}, peak fitting \cite{bib:7} and multi-isotope identification \cite{bib:5, bib:6, bib:8}. Most of the previous studies focus on the multi-isotope identification with sufficient statistics, either for high activity radioactive source or with a long measurement time. In this work, we explore the use of multilayer perceptron (MLP) neural network in environmental radioisotope monitoring.

The application of neural network in spectrum analysis always requires sufficient data for training the neural network. Most of the previous works use measured spectrums to train the neural network, requiring a sufficient number of measurements of standard samples to get a good training dataset. Thus, the application of the neural network is limited by the large amount of work in preparing and measuring the standard samples. Monte Carlo method can be used to provide simulated spectrums to train the neural network. Simulation tools, like Geant4 \cite{bib:9}, can simulate the generation, transportation and absorption of photon in the measurement to give the simulated spectrum. And generating spectrums with known $^{\text{137}}$Cs concentration via Monte Carlo method is expected to greatly save the time required to create a training dataset rather than the measurement of standard samples.

\section{Method} \label{sec:artwork-2}
In the implementation of our MLP-based method, we use Geant4 \cite{bib:9} simulation tool to simulate the $^{\text{137}}$Cs gamma-ray spectrums and combine them with the measured background spectrums to make a training dataset to train the MLP. Then we use the trained MLP to do regression analysis on gamma-ray spectrum to calculated the $^{\text{137}}$Cs concentration. The result is compared to the seawater quality standard to decide the environmental level of the $^{\text{137}}$Cs concentration.

We choose root-mean-square-error (RMSE) to test the regression performance of the MLP neural network. And the ROC area and accuracy are chosen to test the classification ability of this method.

\subsection{Geant4 simulation of Cs-137 gamma-ray spectrum}\label{sec:artwork-2-1}

We use Geant4.10.05 \cite{bib:9} to simulate the energy spectrums of $^{\text{137}}$Cs in the seawater measured by a LaBr$_{\text{3}}$ detector. A model of the monitoring device is built in Geant4, the sketch and details of material and geometry are indicated in Fig. 2. The LaBr$_{\text{3}}$ detector (cylinder with green color) is modelled as a LaBr$_{\text{3}}$ crystal with density of 5.06 g/cm$^{\text{3}}$. The detector box (cylinder with yellow color) containing photomultiplier tube and electronics is simplified to be a cylinder shell made of polyvinyl chloride (PVC) material. The LaBr$_{\text{3}}$ detector and detector box are placed at the center of the device chamber (cylinder with yellow line) filled with seawater (cylinder with blue line), with a 2 cm thick stainless-steel shield (black line) outside. Although the composition of seawater varies with time and location where the equipment is positioned, in our simulation, the sea water is simplified as the mixed solution of NaCl and MgCl$_{\text{2}}$ with a density of 1.05 g/cm$^{\text{3}}$.

\begin{figure}[!htb]
\includegraphics
  [width=1.0\hsize]
  {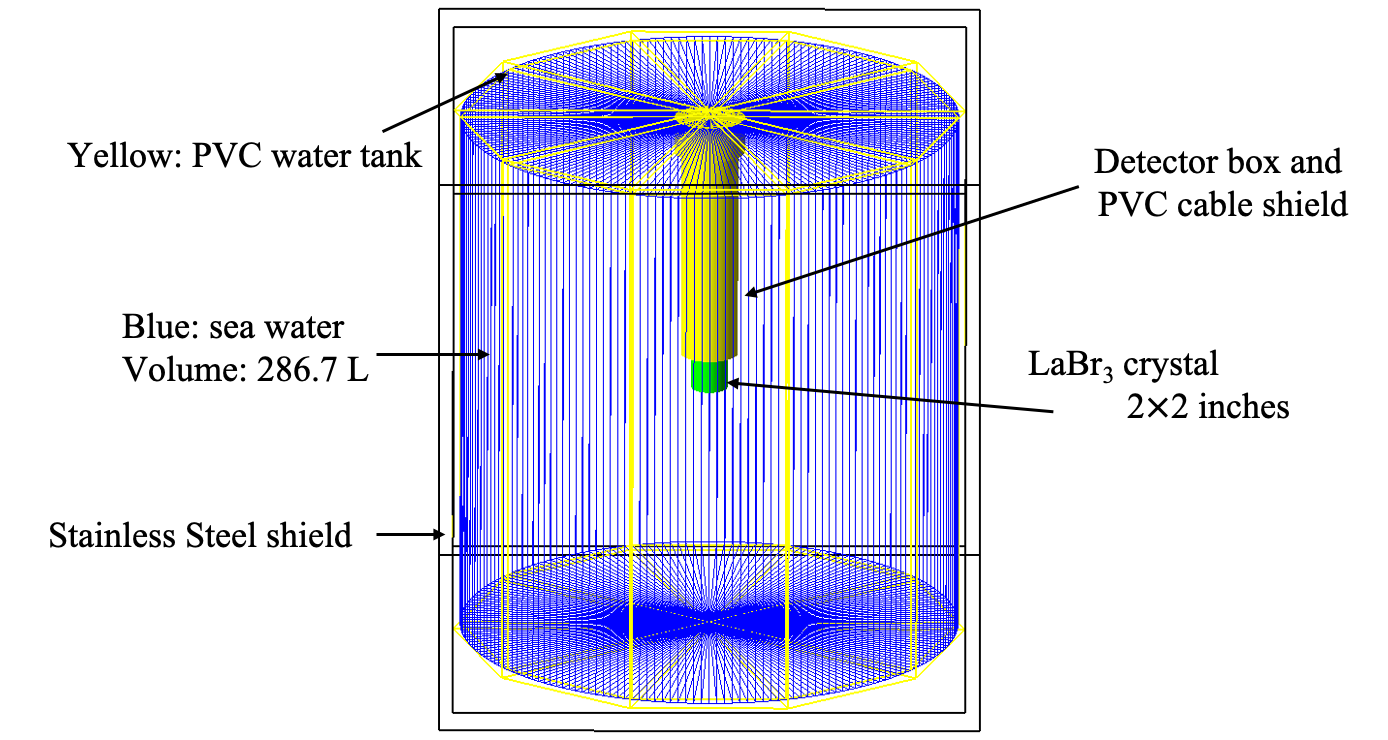}
\caption{Sketch of Geant4 model.}
\label{fig:figure-2}
\end{figure}

Assuming the constant activity within the measurement time, $^{\text{137}}$Cs is homogeneously distributed in the seawater, emitting 662 keV gamma-rays isotropically. The relationship between number of $^{\text{137}}$Cs decay in simulation (N$_{\text{s}}$) and $^{\text{137}}$Cs activity concentration (A$_{\text{Cs}}$, Bq/L) is given by:

\begin{equation}\label{eq-1}
\text{A}_{\text{Cs}} = \frac{\text{N}_{\text{s}}}{\text{V}_{\text{sea}}{\text{I}_{\gamma}}{\text{T}}}.
\end{equation}

In which, V$_{\text{sea}}$ is the volume of seawater (286.73 L), I$_{\gamma}$ is the 662 keV gamma ray intensity (0.86), T stands for the measuring time (3600s). Therefore, the N$_{\text{s}}$ can be calculated using Eq ~\ref{eq-1} for a given $^{\text{137}}$Cs concentration A$_{\text{Cs}}$.

The simulation results do not consider the energy resolution of the LaBr$_{\text{3}}$ detector, we use gaussian broadening in this work to include the energy resolution of the detector. For every deposited energy (E$_{\text{d}}$) in LaBr$_{\text{3}}$ detector, the recorded energy in spectrum (E$_{\text{r}}$) is calculated by random sampling using a gaussian function:

\begin{equation}\label{eq-0}
\text{E}_{\text{r}} = \text{F}_{\text{Gaus}}(\text{E}_{\text{d}},\frac{\text{FWHM}}{\text{2.355}}).
\end{equation}

The F$_{\text{Gaus}}$ is the gaussian function, the full-width-half-maximum (FWHM) is extracted from measured spectrums.

In our simulation, we set different N$_{\text{s}}$ to generate gamma-ray spectrums with $^{\text{137}}$Cs concentration ranging from 0.1 Bq/L to 2.0 Bq/L. 20 groups of $^{\text{137}}$Cs spectrums are simulated, each group consists of 5 independent spectrums with the same $^{\text{137}}$Cs concentration. The $^{\text{137}}$Cs concentration step size between groups is 0.1Bq/L. The minimum and maximum $^{\text{137}}$Cs concentration are 0.1 Bq/L and 2.0 Bq/L.

\subsection{Training and test dataset}\label{sec:artwork-2-2}

The simulated spectrums include the contribution of $^{\text{137}}$Cs, and the measured background spectrums consist of contribution from the background, such as $^{\text{40}}$K in the seawater and intrinsic radioactive isotopes of LaBr$_{\text{3}}$. We combine the simulated spectrums with measured background spectrums to generate the training and test dataset.

We combine 100 simulated signal spectrums with 645 measured background spectrums to generate a dataset of 64500 spectrums for MLP training and test. The combined spectrum is generated by channel-by-channel summation of simulated and measured background spectrums. To focus on the 662 keV peak of $^{\text{137}}$Cs, we select an energy window of 620-710 keV (a total of 30 channels in the combined spectrum) as the input for the following analysis. 

And the total 64500 spectrums are divided for training and test dataset, for each $^{\text{137}}$Cs concentration, 80\% spectrums are randomly selected to form a training dataset, and the rest 20\% spectrums are for an independent test dataset.

\subsection{MLP structure and training process}\label{sec:artwork-2-3}

\begin{figure}[!htb]
\includegraphics
  [width=1.0\hsize]
  {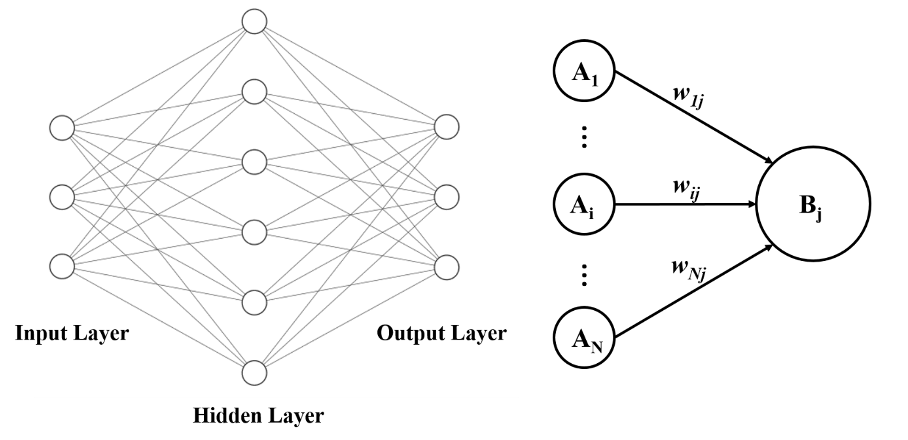}
\caption{An example of MLP (left) and operation of a single neuron (right).}
\label{fig:figure-3}
\end{figure}

The multilayer perceptron (MLP) \cite{bib:12} is one kind of full connecting neural network. An MLP consists of an input layer, series of hidden layers and an output layer, an example of which is shown in Fig.~\ref{fig:figure-3}. The example has an input layer with 3 neurons, one hidden layer with 6 neurons, and an output layer with 3 neurons.

The operation of a single neuron is illustrated in Fig.~\ref{fig:figure-3} (right), in which A$_{i}$ and B$_{j}$ are the outputs of neurons, w$_{ij}$ is the weight between neuron B$_{j}$ and A$_{i}$, the output of neuron is calculated by the neurons of the previous layer: 

\begin{equation}\label{eq-2}
\text{B}_{j} = f(\sum{\omega_{i,j}}\text{A}_{i}).
\end{equation}

Where f is the active function to introduce the nonlinearity to the neural network. 

In this work, we build a predictive model with an input layer with 30 neurons (corresponding to 30 channels in the selected energy window), 10 hidden layers, and an output layer with a single neuron (representing the $^{\text{137}}$Cs concentration).

We choose the Levenberg Marquardt method to train our predictive model using the MatLab neural network tool. In the training, the training dataset (52245 spectrums with known $^{\text{137}}$Cs concentration) is randomly divided: 70\% for training, 15\% for validation, and 15\% for test. The sub-dataset for training and validation parts are used to adjust the weight and hyper-parameters (e.g. learning rate of the model), while the sub-dataset for test part is used to monitor the training process to prevent over-fitting. The epochs of validation step are determined based on the minimum mean squared error (MSE) of validation sub-dataset:

\begin{equation}\label{eq-3}
\text{MSE} = \frac{1}{\text{N}}\sum{(\text{x}-\text{x}_{\text{T}})^2}.
\end{equation}

Where N is the total number of the spectrum used in training, x is the output value of the MLP neural network, and x$_{\text{T}}$ is the actual $^{\text{137}}$Cs concentration of the corresponding spectrum.

After training, the MLP neural network is used to predict the $^{\text{137}}$Cs concentration of an unknown spectrum. And to describe the regression ability, apart from root mean squared error (RMSE) which is the root value of MSE, we also use average relative deviation (D$_{\text{ave}}$):

\begin{equation}\label{eq-4}
\text{D}_{\text{ave}} = \frac{1}{\text{N}}\sum{| \frac{\text{x}-\text{x}_{\text{T}}}{ \text{x}_{\text{T}} } |}.
\end{equation}

\subsection{Spectrum classification}\label{sec:artwork-2-4}

The $^{\text{137}}$Cs concentration should be warned if it exceeds a limit. The spectrum is classified by comparing the output of MLP to a threshold. If the output value is below or over the threshold, the spectrum is classified to be normal and abnormal, respectively. In this study, we set the threshold equal to 0.7 Bq/L according to the China National Standard of seawater quality \cite{bib:11}.

However, due to the uncertainty of output value, it is possible to misjudge some normal spectrums (so-called false positive mistakes) and miss some abnormal spectrums (so-called true negative mistakes). We use the receiver operating characteristic (ROC) curve and maximum accuracy to describe the classification ability.

The ROC curve and maximum accuracy are calculated by scanning the threshold form the minimum to the maximum output value, for each threshold, the false positive rate (FP), the true positive rate (TP) and the accuracy (AC) are calculated by:

\begin{equation}\label{eq-5}
\text{FP}=\frac{\text{N}_{\text{FP}}}{\text{N}_{\text{F}}}, \quad \text{TP}=\frac{\text{N}_{\text{TP}}}{\text{N}_{\text{P}}}, \quad
\text{AC}=\frac{\text{N}_{\text{C}}}{\text{N}_{\text{T}}}.
\end{equation}

In which, \(\text{N}_{\text{FP}}\) is the number of spectrums with actual $^{\text{137}}$Cs concentration < the limit and output value over the threshold; \(\text{N}_{\text{F}}\) is the number of spectrums with actual $^{\text{137}}$Cs concentration < the limit; \(\text{N}_{\text{TP}}\) is the number of spectrums with actual $^{\text{137}}$Cs concentration > the limit and output value over the threshold; \(\text{N}_{\text{P}}\) is the number of spectrums with actual $^{\text{137}}$Cs concentration > the limit. \(\text{N}_{\text{C}}\) is the number of spectrums with the right classification, and \(\text{N}_{\text{T}}\) is the total number of spectrums.

The maximum accuracy is the highest AC obtained in the scan and the ROC curve \cite{bib:13} is the curves of TP versus FP.

The classification ability is often considered as good when a classifier could achieve a high true positive rate (TP) at a low false positive rate (FP). Therefore, the ROC-area (integral of the ROC curve) can be used to demonstrate the classification ability \cite{bib:13}.

\section{RESULTS and DISCUSSION} \label{sec:artwork-3}
\subsection{Result of Geant4 simulation and MLP training}\label{sec:artwork-3-1}

\begin{figure}[!htb]
\includegraphics
  [width=1.0\hsize]
  {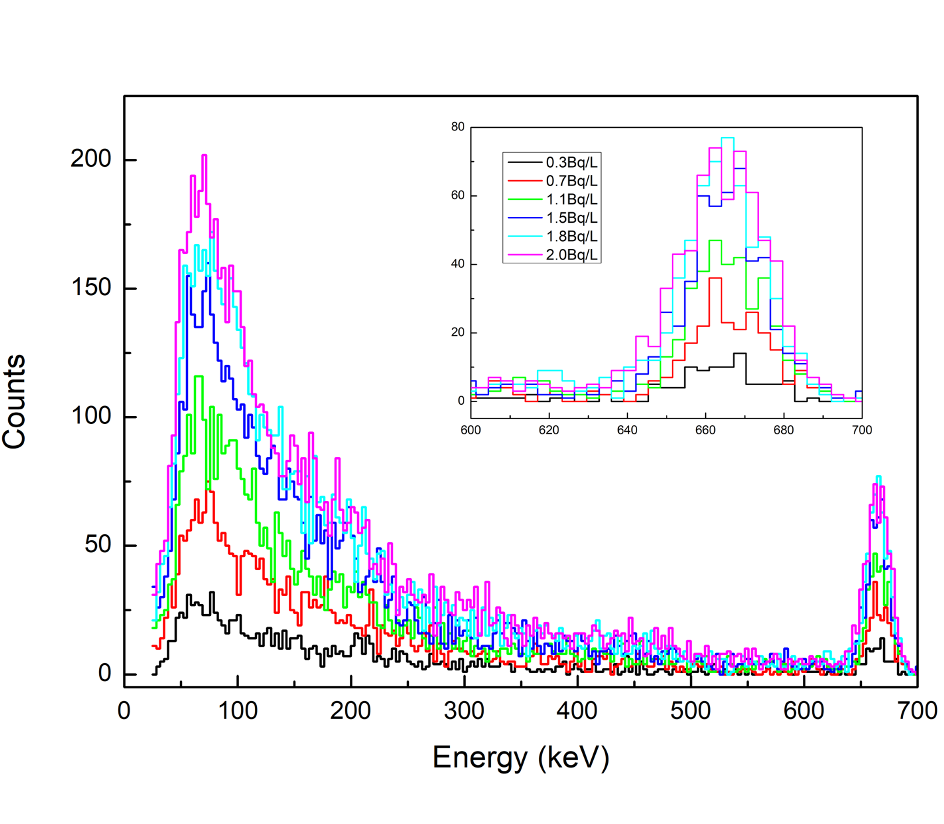}
\caption{Simulated single gamma-ray spectrums with different $^{\text{137}}$Cs concentrations.}
\label{fig:figure-4}
\end{figure}

100 $^{\text{137}}$Cs gamma-ray spectrums with 0.1-2.0 Bq/L $^{\text{137}}$Cs concentration are simulated using Geant4, 6 spectrums are randomly selected and shown in Fig.~\ref{fig:figure-4}. The 662keV full energy peak of $^{\text{137}}$Cs is clearly seen in Fig.~\ref{fig:figure-4}, however, due to the poor statistics, the other spectrum features like the Compton edge and back scatting peak are not visible.

\begin{figure}[!htb]
\includegraphics
  [width=1.0\hsize]
  {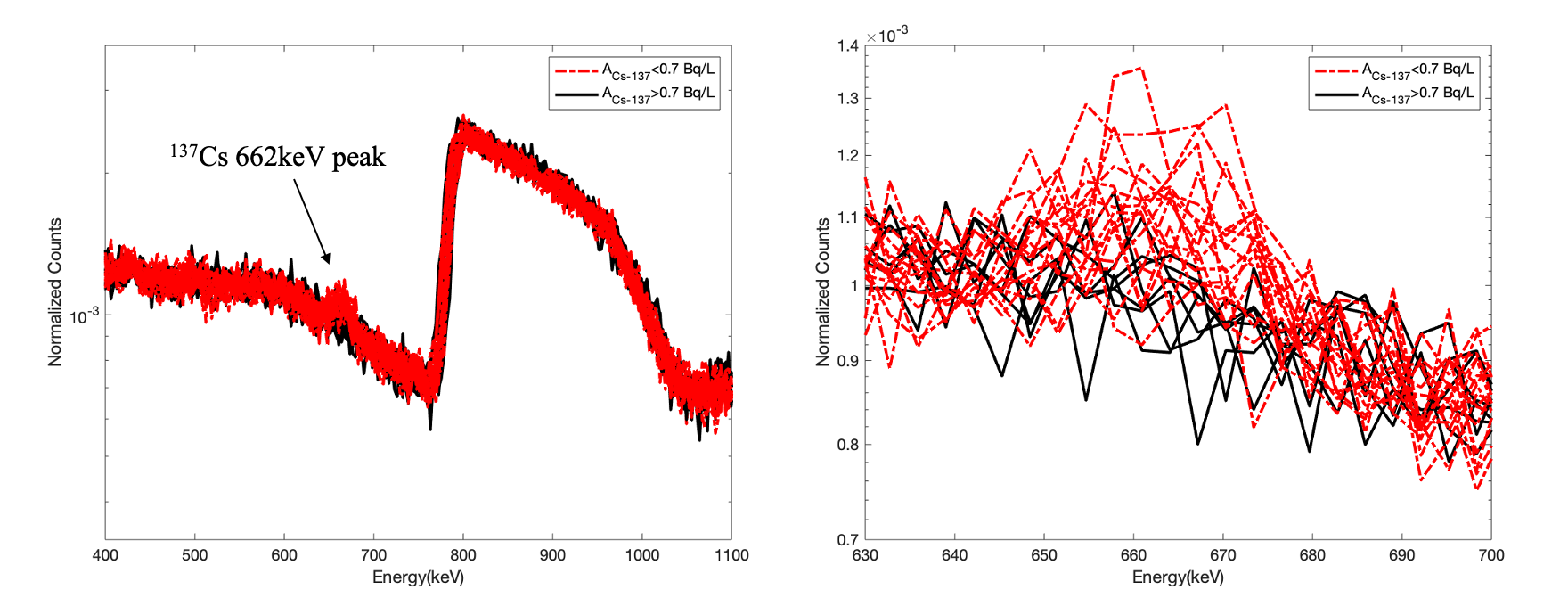}
\caption{Combined $^{\text{137}}$Cs spectrums (normalized) around 662keV, randomly selected.}
\label{fig:figure-5}
\end{figure}

The 100 $^{\text{137}}$Cs spectrums are combined with 654 background spectrums, and the combined spectrums are randomly selected and shown in Fig.~\ref{fig:figure-5}. The peak in 700-1100 keV energy region is contributed by the beta- decay of intrinsic radioactive isotope $^{\text{138}}$La.

\begin{figure}[!htb]
\includegraphics
  [width=1.0\hsize]
  {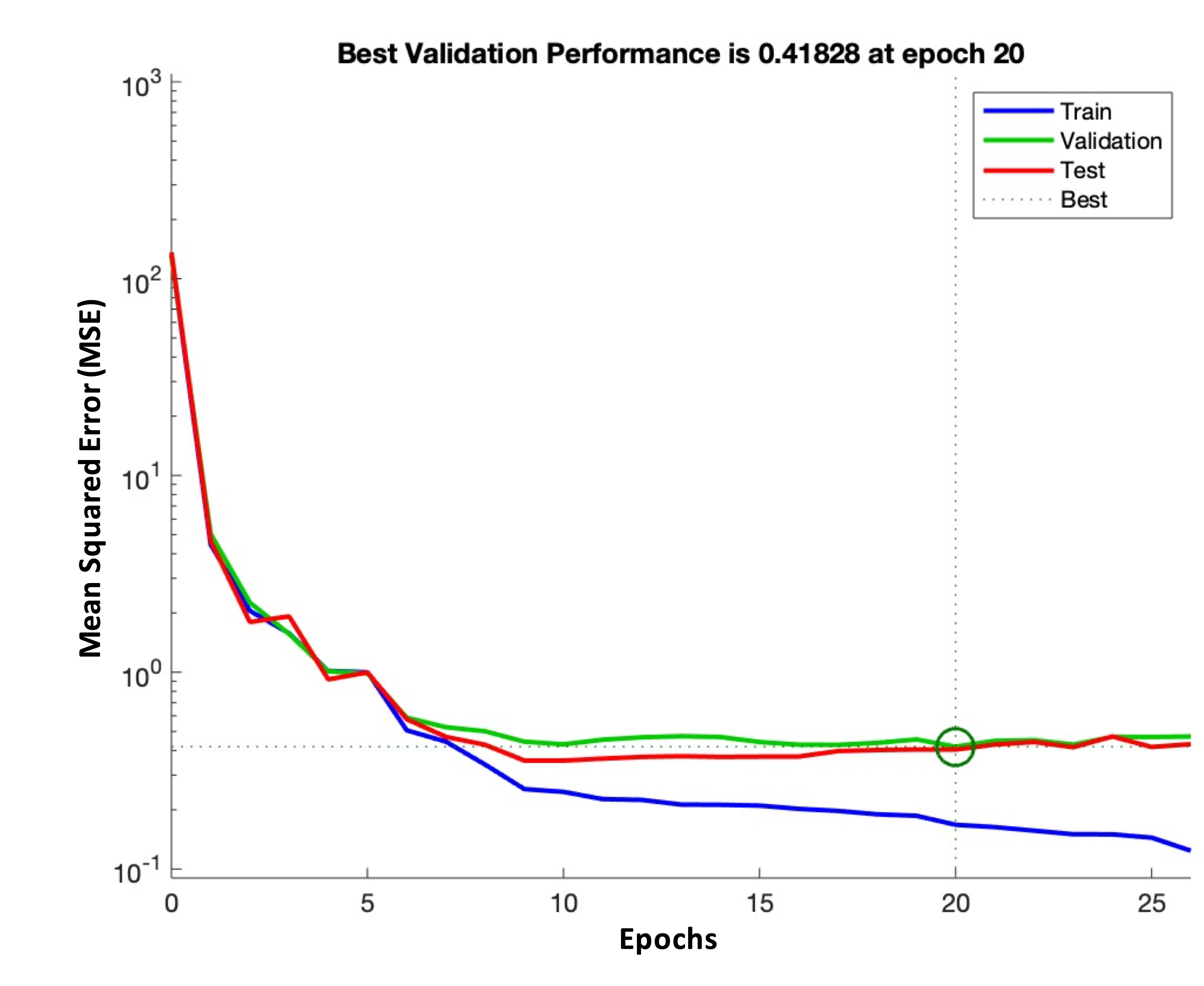}
\caption{Training process of MLP model.}
\label{fig:figure-6}
\end{figure}

The training process of MLP is shown in Fig.~\ref{fig:figure-6}, the training stops at 26$^{\text{th}}$ epoch, and the best result is found at the 20$^{\text{th}}$ epoch (marked with green circle) where the minimum MSE of validation sub-dataset is observed.

\subsection{Test results of MLP-based method and comparison with traditional method:}\label{sec:artwork-3-2}

\begin{figure}[!htb]
\includegraphics
  [width=1.0\hsize]
  {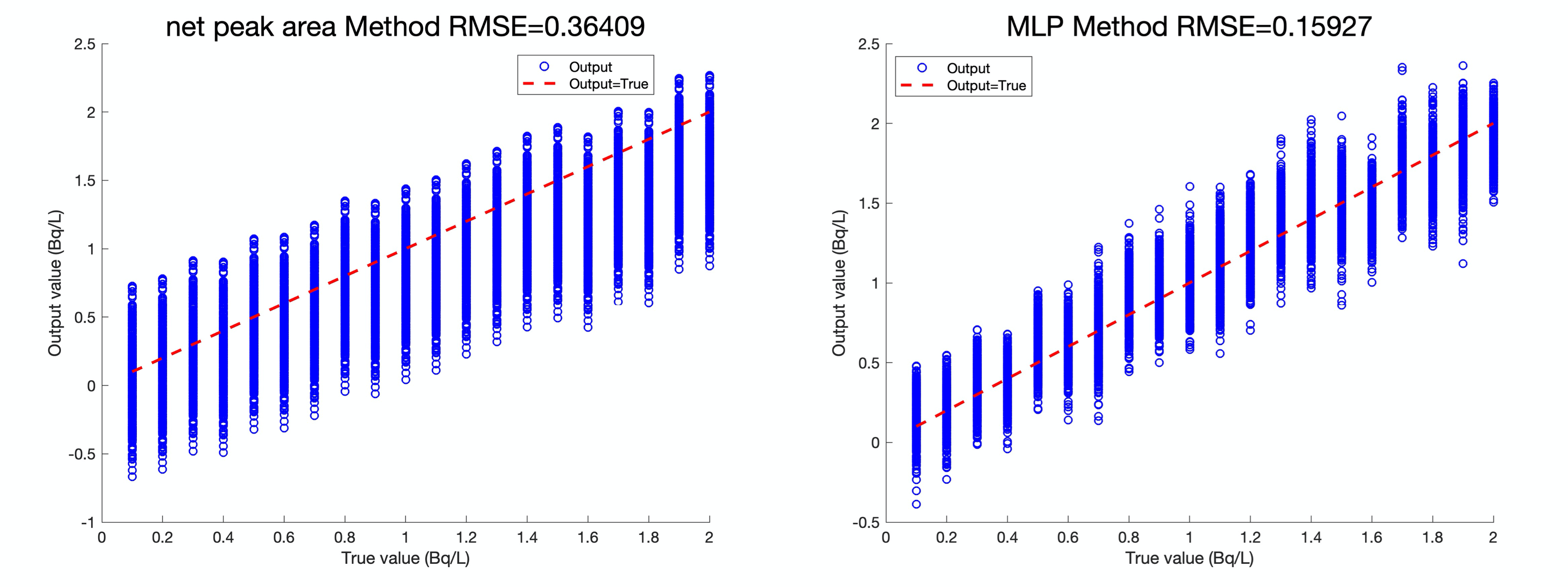}
\caption{The output of net peak area method (left) and the MLP-based method (right). The red dash lines represent the ideal case that output value is equal to actual value.}
\label{fig:figure-7}
\end{figure}

We apply both the trained MLP and traditional net peak area (NPA) method to the test dataset (12900 spectrums with known $^{\text{137}}$Cs concentration). Fig.~\ref{fig:figure-7} shows the outputs of MLP and NPA in the test dataset, in which, the root mean squared error (RMSE) is the root of MSE (Eq.~\ref{eq-3}). Compared to the traditional net peak area method, the MLP-based method achieves a 56.3\% improvement in RMSE.

\begin{figure}[!htb]
\includegraphics
  [width=1.0\hsize]
  {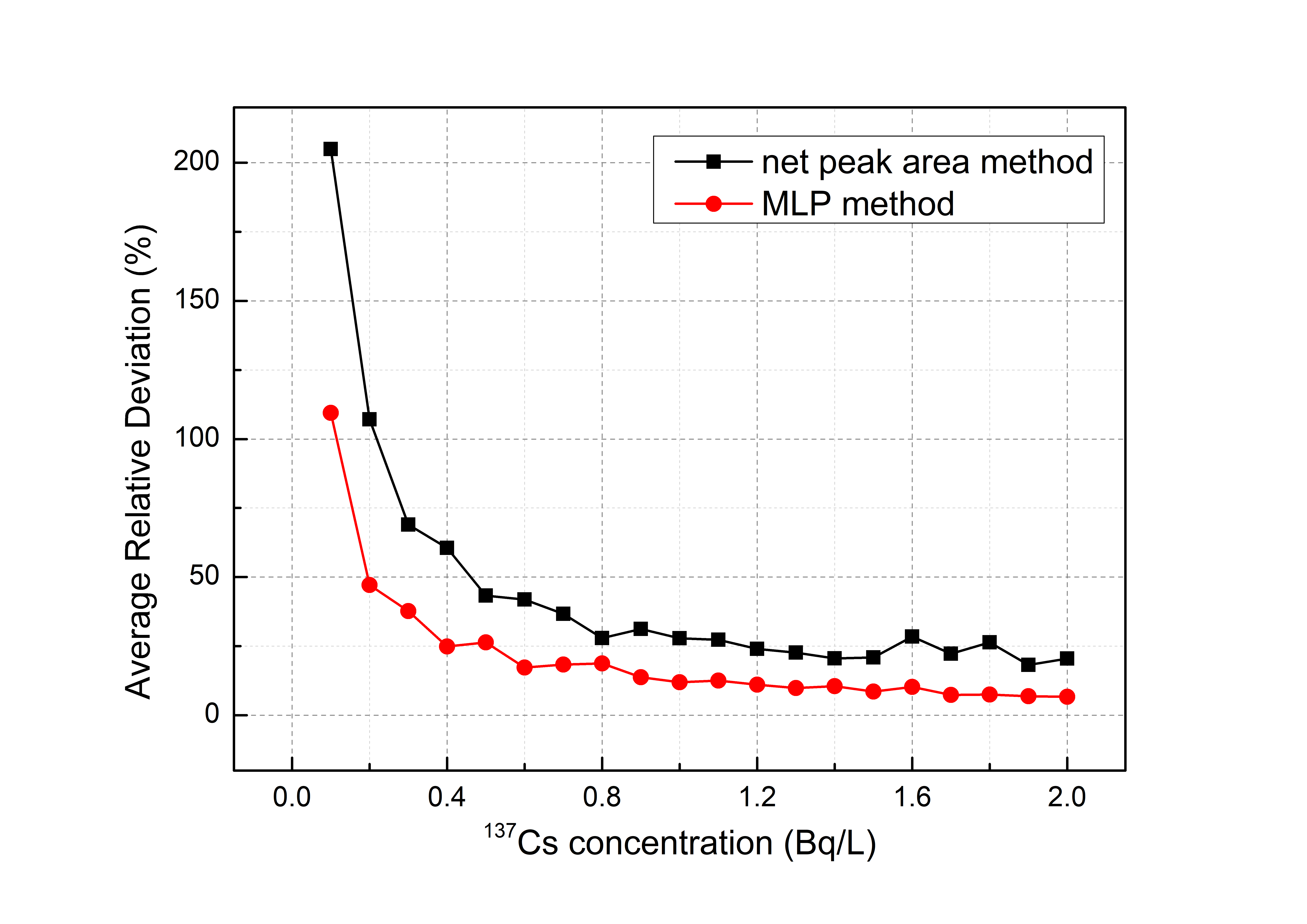}
\caption{Average relative deviation as a function of $^{\text{137}}$Cs concentration for the traditional net peak area method and the MLP-based MLP method.}
\label{fig:figure-8}
\end{figure}

Fig.~\ref{fig:figure-8} shows the average relative deviation as a function of $^{\text{137}}$Cs concentration. For both methods, the average relative deviation decreases as the $^{\text{137}}$Cs concentration increase, this is because when the $^{\text{137}}$Cs signal is small, the spectrum is dominated by the statistical fluctuation of the background. But our new method maintains a lower average relative deviation than the traditional method in Fig.~\ref{fig:figure-8}.

\begin{figure}[!htb]
\includegraphics
  [width=1.0\hsize]
  {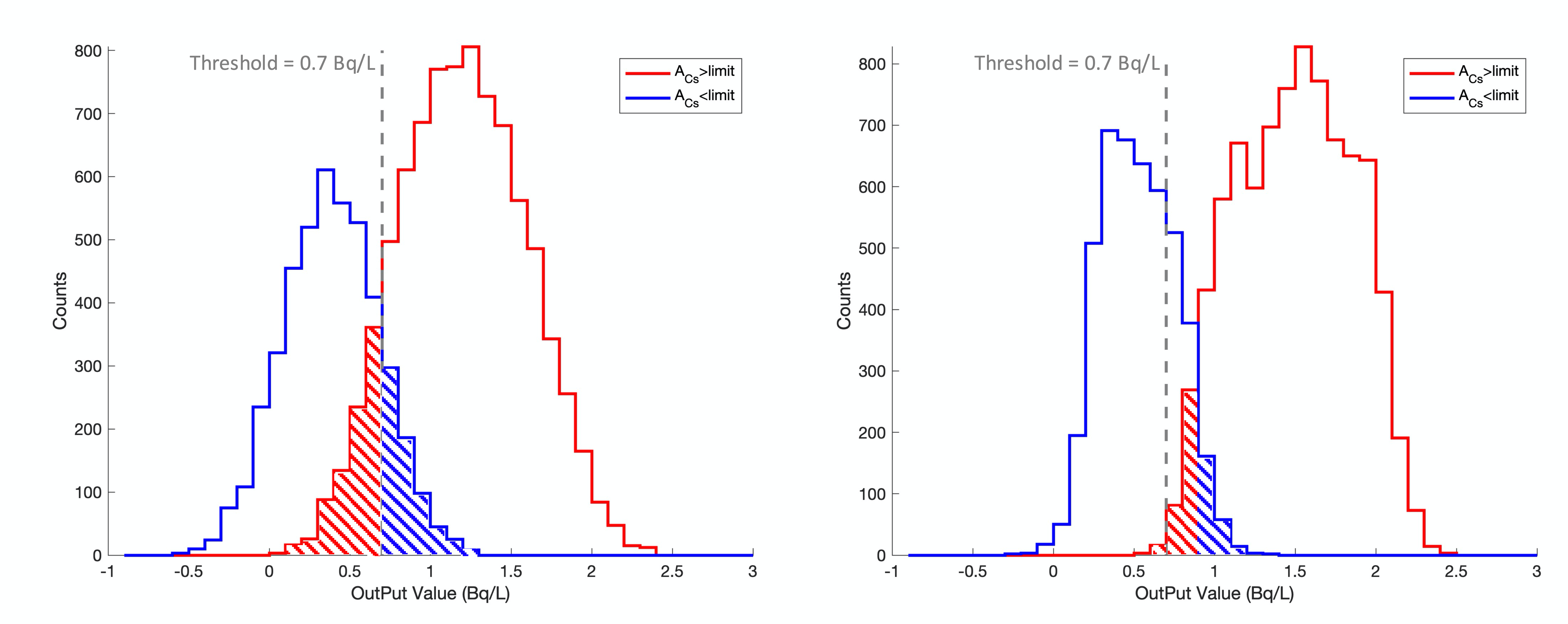}
\caption{Output value distribution of the traditional net peak area method (left) and the MLP-based method (right). The distribution overlap of output value with actual $^{\text{137}}$Cs concentration exceeds and below the limit are showed in red and blue shadow regions. The gray dash line is the threshold in spectrum classification.}
\label{fig:figure-9}
\end{figure}

The distributions of output value are showed in Fig.~\ref{fig:figure-9} for both the traditional NPA method and the MLP-based method. The output value is divided into 2 group: one with actual $^{\text{137}}$Cs concentration below the limit and the other with actual $^{\text{137}}$Cs concentration over the limit. Due to the calculation error in $^{\text{137}}$Cs concentration, there is an overlap between the distribution of the 2 groups of output. Since the MLP-based method has a lower calculation uncertainty of $^{\text{137}}$Cs concentration, the overlap region of the MLP-based method is smaller than that of the traditional NPA method. And when the threshold (gray dash line in Fig.~\ref{fig:figure-9}) is equal to the limit, our new method identifies 99.8\% of spectrums with actual $^{\text{137}}$Cs concentration over the limit, while the traditional NPA method only identifies 89.7\% of them.

\begin{figure}[!htb]
\includegraphics
  [width=1.0\hsize]
  {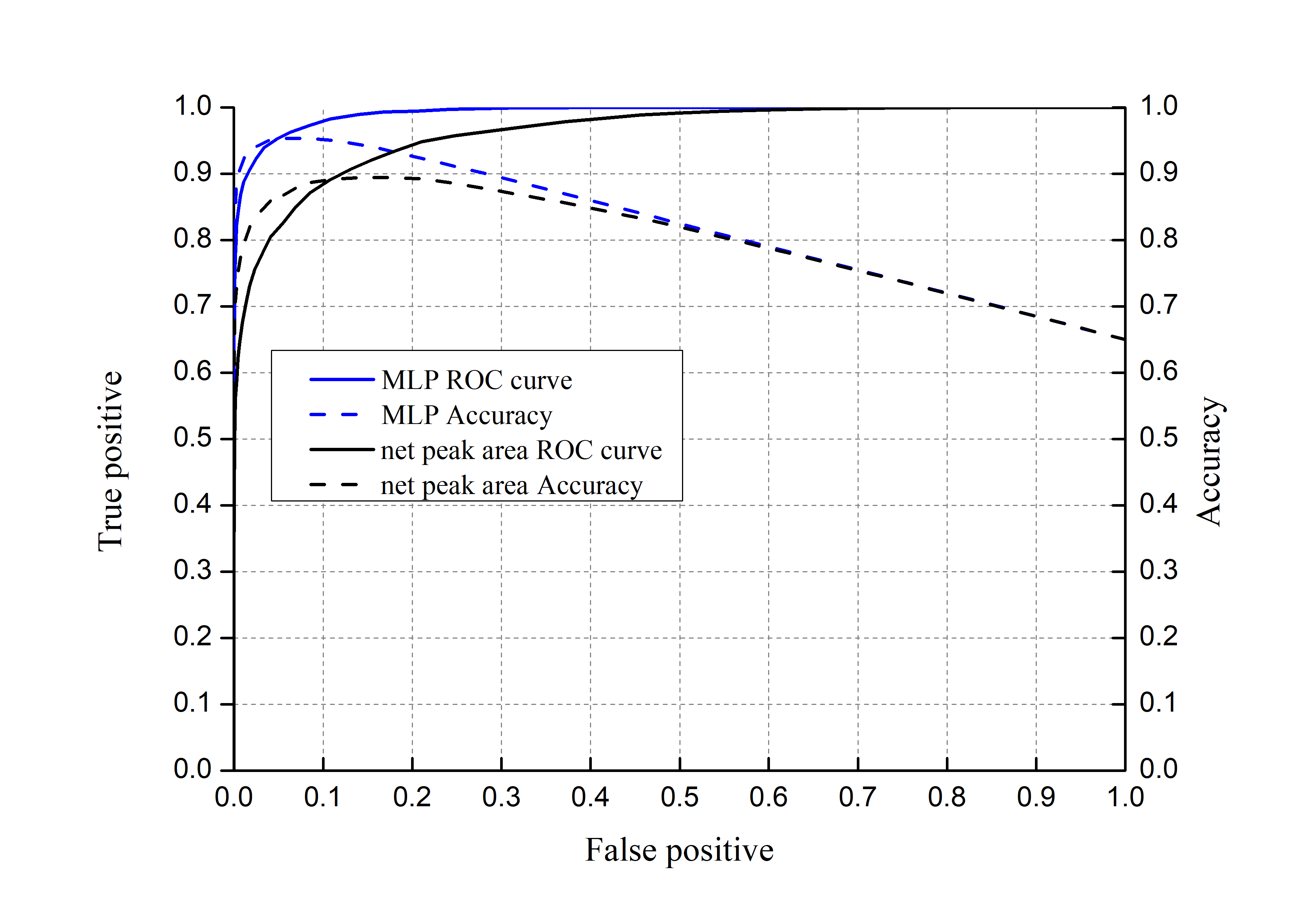}
\caption{ROC curve of MLP-based method in testing set (blue line) and net peak area method in testing set (black line); the blue and black dash line are the accuracy of the two methods corresponding to each TP and FP.}
\label{fig:figure-10}
\end{figure}

The ROC-area and accuracy for both methods are listed in Table~\ref{tab:table-1}. Comparing with the NPA method, the MLP-based method obtains a 3.8\% improvement in ROC area and a 9\% improvement in accuracy. The ROC and accuracy curves are shown in Fig.~\ref{fig:figure-10}, the higher true positive rate and accuracy at the same false positive rate for the MLP-based method indicate a better classification performance.

\begin{table}[htbp]
\centering
\caption{\label{tab:table-1} {Classification performance of MLP-based method and net peak area method.}}
\smallskip
\begin{tabular}{|lr|c|}
\hline
Method&ROC-area&Accuracy(threshold=0.7Bq/L)\\
\hline
MLP-based method & 0.9932	& 0.9445\\
Net peak area method & 0.9567 &	0.8665 \\
\hline
\end{tabular}
\end{table}

Despite of the improvement on calculation and classification performance of low-level $^{\text{137}}$Cs, the analysis results are strongly depended on the training dataset. In this work, we assumed that the performance of the detection system and the background are stable. In practice, the shift of detector performance (linearity and resolution of energy) and variations of environmental backgrounds may require the reconstruction of training dataset and re-training of the predictive model. And due to the ‘black box’ nature of the neural network, the analysis results are more difficult to explain compared to those of the traditional net peak area method.

\section{Impact assessment of the assumption of Cs-137 concentration in training dataset} \label{sec:artwork-4}

In the training dataset, we assume that there is no $^{\text{137}}$Cs in the background spectrums measured by the monitoring device in its sea trail. To validated the assumption, some seawater samples are taken and analyzed by the radiochemistry method alone with the measurement. The results show the $^{\text{137}}$Cs concentrations of the seawater samples are in range of 1 to 3 mBq/L which are 2 orders lower than the $^{\text{137}}$Cs concentration limit (0.7 Bq/L), indicates the $^{\text{137}}$Cs in measured background spectrum is negligible.

However, since the limited number of seawater samples, not every background spectrum has a seawater sample for radiochemistry analysis. Despite there is no observable 662keV $^{\text{137}}$Cs gamma-ray peak in all background spectrums, it is theoretically possible that the measured background spectrums without radiochemistry analysis have $^{\text{137}}$Cs concentration near the detection limit of the device (0.48 Bq/L), which will lead to a violation of the 0 Bq/L $^{\text{137}}$Cs concentration assumption of background spectrums. Therefore, we randomly set the $^{\text{137}}$Cs concentration in background spectrums to evaluate the impact of this potential violation on the performance of the MLP-based method.

The $^{\text{137}}$Cs concentration of 645 measured background spectrums are randomly sampled from 0 to 0.48 Bq/L. Then, we update the training and test dataset and retrain our predictive model. Afterwards, we test our model alone with the traditional net peak area method in the test dataset and record the RMSE of both methods for comparison.

\begin{figure}[!htb]
\includegraphics
  [width=1.0\hsize]
  {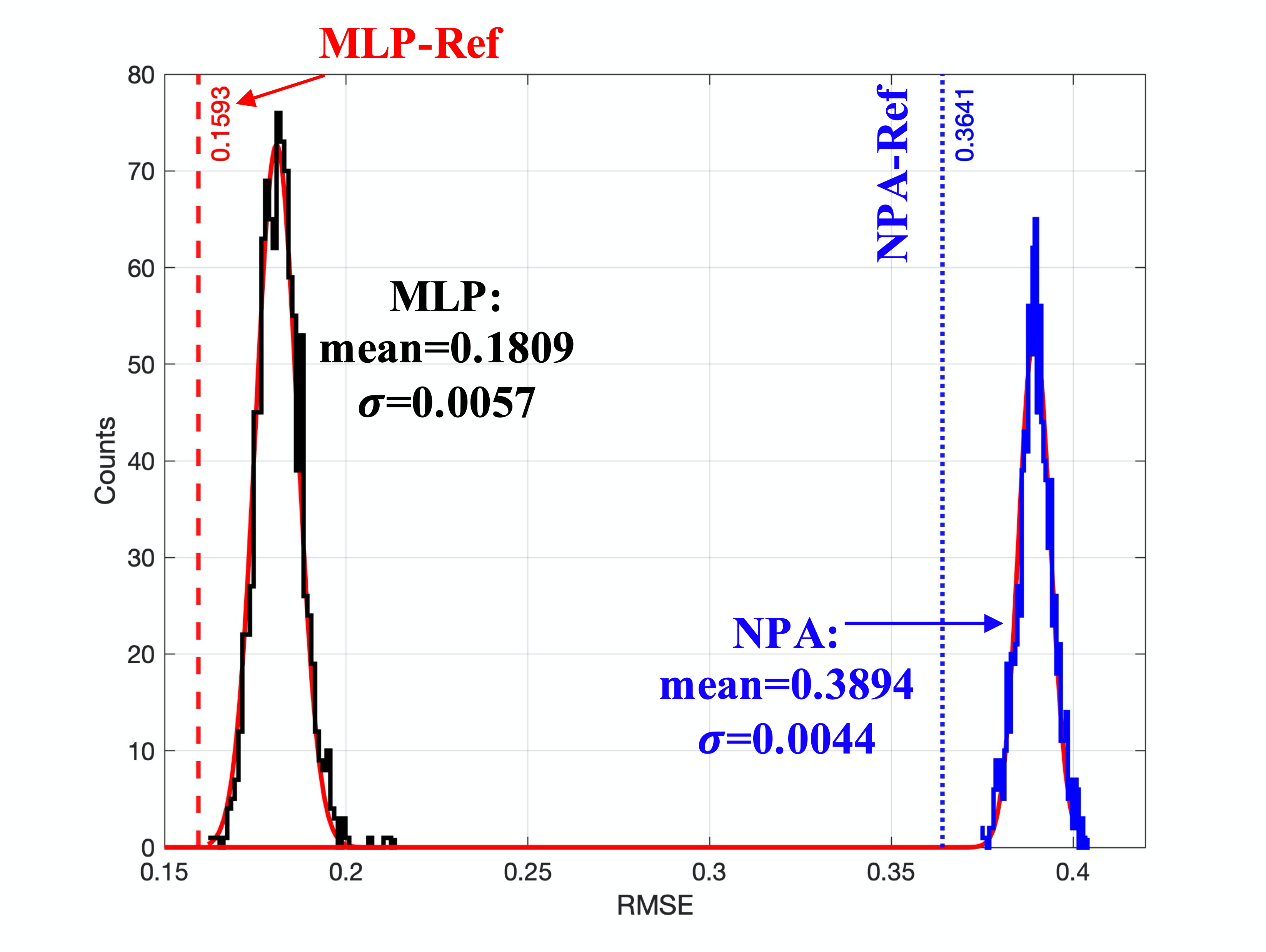}
\caption{RMSE distribution of MLP (black line) and NPA (blue line) method; The red solid lines are the gaussian-fit curves of RMSE distribution, the red dash lines (MLP Ref) and blue dotted lines (NPA Ref) are the RMSE of MLP and net peak method under the assumption of $^{\text{137}}$Cs concentration of 0 Bq/L.}
\label{fig:figure-11}
\end{figure}

We repeat the aforementioned procedure for 1000 times, and get the distribution of the RMSE. The distribution is gaussian fitted and the mean and standard deviations of Gaussian function are extracted to represent the average performance and stability of both methods. The results are presented in Fig.~\ref{fig:figure-11}.

In Fig.~\ref{fig:figure-11}, the RMSE of both methods have increased compared to the result under the assumption of null $^{\text{137}}$Cs in measured background spectrums. A probable reason is that when we randomly set the $^{\text{137}}$Cs concentrations in range of 0 to 0.48 Bq/L, we overestimate the fluctuation of the $^{\text{137}}$Cs concentrations in the background spectrums, which deteriorates the physical connection between the spectrums and the actual $^{\text{137}}$Cs concentration and leads to the deterioration of RMSE. Nevertheless, the performance of MLP is still better than that of the traditional net peak area method. And the small standard deviation of our new method, only 3\% of the mean value, indicates that the new method has a good stability under the fluctuation of $^{\text{137}}$Cs concentration in background spectrums.

\section{Conclusion}

In this work, we propose a method utilizing multilayer perceptron (MLP) to analyze gamma-ray spectrums measured by a marine radioisotope monitoring device. Combining the signal spectrums simulated by Geant4 with the measured spectrums, we avoid burden of standard sample measurements while maintaining the quality of training dataset. And compared to the traditional net peak area method, the test results of RMSE and average relative deviation show that the MLP-based method improve the precision of $^{\text{137}}$Cs concentration calculation and the results of classification accuracy and ROC-area show a better spectrum classification ability. The proposed method is proved to be a suitable and stable method for in-situ gamma-ray spectrometry in seawater to monitor artificial radionuclides. For future work, we will conduct a series of standard sample measurement to improve the simulation process and achieve a better network performance.

In addition, the MLP-based method can also be used to monitor other gamma-ray radionuclides via LaBr$_{\text{3}}$ or other types of radiation detectors, like NaI or HPGe.


\acknowledgments

This work is supported by the Tsinghua University Initiative Scientific Research Program.



\end{document}